\documentclass[12pt,preprint]{aastex}

\slugcomment{To appear in ApJ., 2002}

\shorttitle{MICROLENSING ON BROAD EMISSION LINES}

\begin{document}

\title{The Influence of Gravitational Microlensing \\
    on the Broad Emission Lines of Quasars}

\author{C. Abajas\altaffilmark{1}, E. Mediavilla\altaffilmark{1}, J. A.
Mu\~noz\altaffilmark{1},
L. \v C. Popovi\'c\altaffilmark{2}, and A. Oscoz\altaffilmark{1}}
\email{abajas@ll.iac.es, emg@ll.iac.es, jmunoz@ll.iac.es, \\
 lpopovic@aob.aob.bg.ac.yu, aoscoz@ll.iac.es}
 
\altaffiltext{1}{Instituto de Astrof\'\i sica de Canarias, E-38205 La Laguna, Tenerife,
Spain}
\altaffiltext{2}{Astronomical Observatory, Volgina 7, 11160 Belgrade,
Serbia, Yugoslavia}

\begin{abstract}

We discuss the effects of microlensing on the broad emission lines (BELs) of
QSOs in the light  of recent determinations of the size of the broad line
region (BLR) and its scaling with  QSO luminosity. Microlensing by
star-sized objects can produce significant  amplifications in the BEL of some
multiple-imaged QSOs, and could be very relevant  for high-ionization lines. We
have identified a group of ten gravitational lens systems ($\sim$ 30\% of the
selected sample) in which microlensing could be observed. Using standard kinematic
models for AGNs, we  have studied the changes induced in the line profile by a
microlens located at different  positions with respect to the center of the
BLR. We found that microlensing could produce important effects such as
the relative enhancement of different parts of the line profile or the 
displacement of the peak of the line.
The study of BEL profiles of different ionization in a microlensed QSO
image could be an alternative method for probing the BLR structure and size. 

\end{abstract}


\section{INTRODUCTION}

The change in the continuum flux of quasars by stars or compact objects
in intervening  galaxies (gravitational microlensing) is now a
well-established observational phenomenon. 
Several studies have attempted to resolve the structure of
the region generating the  optical and UV continuum by using the
microlensing effect as a gravitational telescope  (see Yonehara 1999, and
references therein). However, it has  usually been assumed that  the size
of the region generating the broad emission lines (BELs) of quasars is
too large  (0.1--1 pc within the framework of the standard model, Rees
1984)  to be substantially affected  by stellar-mass lenses. According to
previous studies (Nemiroff 1988; Schneider \& Wambsganss  1990), by
comparing one line in different components of a multiple-image quasar,
we  would be able to detect only fractional deviations of the lensed from
the unlensed profile.  Even using optimistic estimates for the microlens
masses, these deviations would  amount to a modest 10--30\% range.
However,  new results concerning the BLR structure and kinematics are
relevant to these studies, and could  substantially change the common
view about the expected influence of microlensing events on the BEL
profiles. A recent study (Wandel, Peterson, \& Malkan 1999) that
compares  the BLR size determinations obtained using both the
reverberation and the photoionization  techniques for a sample of 19 low
luminosity AGN (mainly Seyferts) inferred a size in the range of a few
light days  to a few light weeks. For AGN of greater luminosity (QSOs)
Kaspi et al. (2000) derived sizes from the Balmer lines in the range from 13
to 514 light days, finding a global scaling of the BLR size as a function of the
intrinsic luminosity, $r_{\rm BLR}\propto L^{0.7}$.

The purpose of this paper is to revisit the influence of microlensing in
the BEL  by incorporating these new results.  In \S 2 we estimate the
possibility of microlensing on the BLR of different known multiple-image
quasars. In \S 3 we use simple, but not kinematically unrealistic, models
of  BLRs to study qualitatively the effects that microlensing can produce
on the line profiles.  In \S 4 the observational implications and
applications will be discussed. Finally, the main conclusions are
summarized in \S 5.

\section{ESTIMATES FOR SOME KNOWN SYSTEMS OF MULTIPLE-IMAGE
QSOs}

Only extended objects of sizes comparable to or smaller than the Einstein
radius associated with the gravitational lens will experience  appreciable
amplifications (e.g., Schneider, Ehlers, \& Falco 1992).  Thus, in the
framework of the AGN standard model where the BLR is supposed to have a
radius in the 0.1--1 pc range, only massive  deflectors could give rise
to significant amplifications. However, recent results from the MACHO 
project indicate that the most likely microdeflector masses in the
Galactic halo are in the range 0.15--0.9 $M_{\sun}$ (Alcock et al.
2000b). In the Galactic bulge the microdeflector masses are in good
agreement with normal-mass stars (Alcock et al. 2000a). Estimates from
the lightcurve of Q~2237+0305 (Wyithe, Webster, \& Turner 2000) are also in
reasonable
agreement  with these values. Consequently, significant amplifications of
the BEL would not be expected  according to the standard model. This
result has been pointed out by other  authors (Nemiroff 1988; Schneider
\& Wambsganss 1990), even if they  were somewhat optimistic concerning
the distribution of the microlens mass adopted. 

However, recent research on the BEL seems to indicate that the BLR could be
substantially smaller  than expected in the standard model (Wandel et al.
1999).  Based on the idea that `the continuum/emission line 
cross-correlation function measures the responsivity-weighted radius of
the BLR (Koratkar  \& Gaskell, 1991)', Wandel et al. (1999) have
obtained reliable size measurements.  The sizes obtained using this
technique (reverberation mapping) are consistent with the  substantially
more extensive but less accurate measurements inferred from the
photoionization  method. The results summarized by Wandel et al. (1999)
exhibited a large scatter  (from 1.4 to 107 days)  which could be
attributed to: i) the different size/structure of the BLR  in different
objects, and ii) the different sizes of the regions associated  with
emission lines  of different degrees of ionization. The latter
possibility has been neatly exemplified in the case  of NGC 5548, in
which luminosity-weighted radii ranging from $\sim$ 2.5 days for the He~{\sc ii}
$\lambda$1640 line to  the 28 days corresponding to the [C~{\sc iii}]
$\lambda$1909 line have been found (Peterson \& Wandel 1999). NGC 5548
and the other objects in the sample of Wandel et al. (1999) are
AGN of relatively low luminosity. However, there is a range of
luminosities among the lensed QSOs. To take this fact into account we 
use the results from Kaspi et al. (2000), who found a dependence
between the BLR size and the intrinsic luminosity of the AGN,
$r_{\rm BLR}\propto L^{0.7}$. 

\subsection{Search for candidates}

For a typical lens configuration ($z_{\rm l} =0.5$ and $z_{\rm s} =2$), the projected
Einstein radius of the microlens on the source is $\eta_0\sim 20
(M/M_{\odot})^{(1/2)}$ light-days (for a $\Omega=0.3$ flat cosmology and
$H_0=70$ km s$^{-1}$ Mpc$^{-1}$). Objects with $r_{\rm BLR}\lesssim \eta_0$ are
significantly affected by microlensing. So the relevant question is whether in
the sample of the $\sim 70$ known gravitational lens systems there exist
some lensed quasars with BLR radius $r_{\rm BLR} \lesssim  \eta_0$. To
answer this question we can apply the Kaspi relationship
$r_{\rm BLR}\propto L^{0.7}$ using NGC 5548 as a reference object. We adopt two
values for the BLR size of NGC 5548 in order to account for its
stratification: 21.2 light-days (Balmer lines, Kaspi et al. 2000) and 2.5
light-days (high ionization, He~{\sc ii} line, Peterson \& Wandel 1999). One can
straightforwardly verify  that for a $z=2$ (1) quasar with $m_V = 24.3$
(22.5) would have the same $r_{\rm BLR}$ as NGC 5548, i.e., the microlensing
on the BLR would be quite pronounced. Multiple-image objects of this
apparent magnitude can be detected and, in fact, there are
several examples among the currently known gravitational lens systems.  

\begin{figure}
\figurenum{1}
\plotone{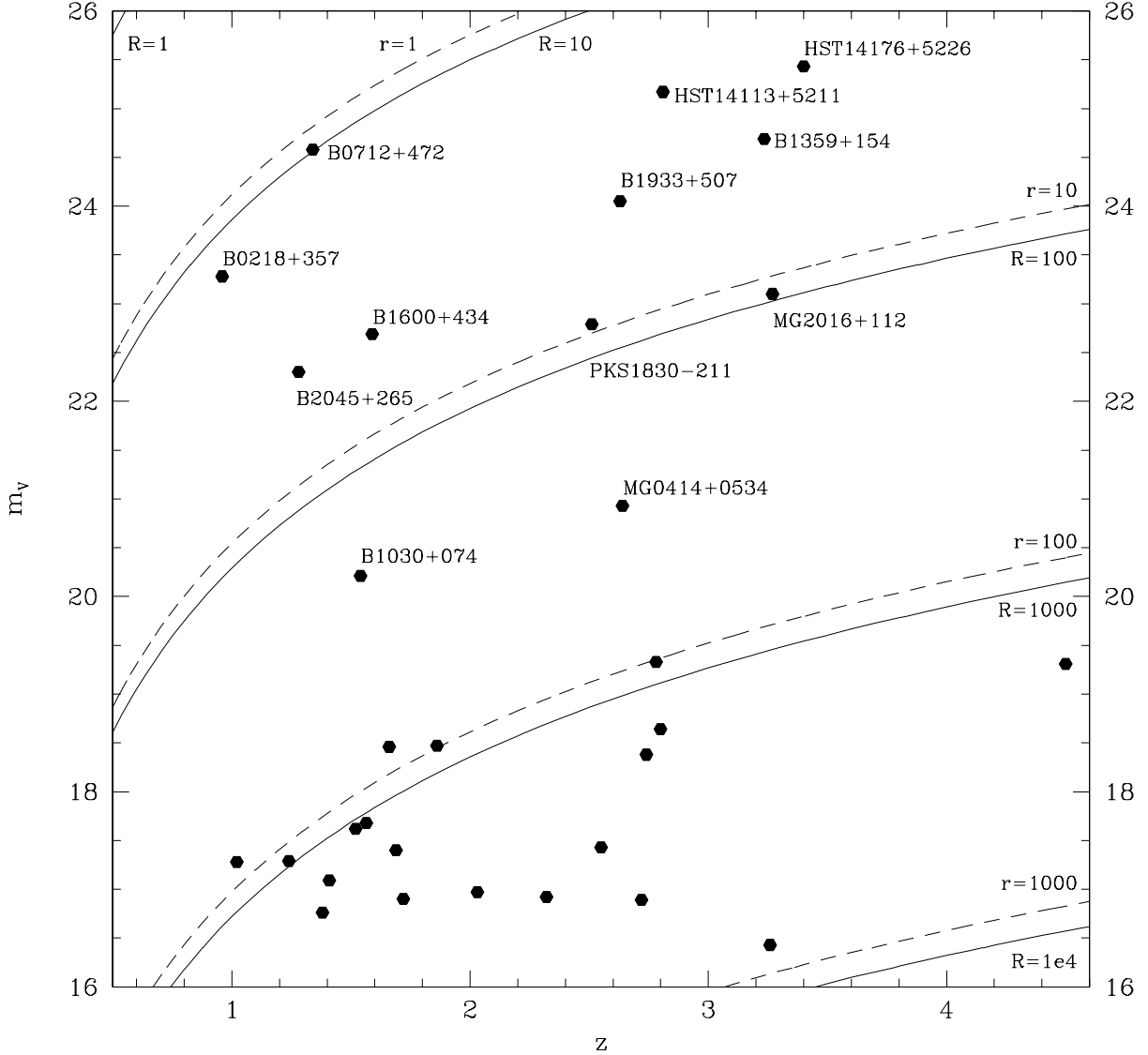}
\figcaption{Contour plots of the BLR radius (in light-days) as a function of  source redshift
and apparent magnitude using
both reference values for the BLR of NGC 5548 ($R$ when $r_{\rm
BLR}({\rm NGC\ 5548})=21.2$ light-days and r 
when $r_{\rm BLR}({\rm NGC\ 5548})=2.5$ light-days) and the Kaspi et al.
relationship $r_{\rm BLR}\propto L^{0.7}$ 
(for an $\Omega=0.3$ flat cosmology and $H_0=70$ km s$^{-1}$ Mpc$^{-1}$). 
Points represent a sample of 31 QSOs whose redshift--magnitude values have been
observed. No extinction or amplification corrections are taken into account.\label{fig1}}

\end{figure}

We can refine this rough estimate by taking into account the redshift and
apparent magnitude of each lensed object. With this aim, in Figure
\ref{fig1} we show contour plots of the BLR radius as a function of the
source redshift and apparent magnitude using both reference values for
the BLR of NGC 5548 and the $r_{\rm BLR}\propto L^{0.7}$ law. We have also
included in Fig. \ref{fig1} the observed redshift--magnitude values
corresponding to a sample of 31 QSOs selected from the $\sim 70$ known
gravitational lenses with the criteria of having the lens
and source  redshifts and the optical magnitude well determined. 
We have used the $V$ magnitude of the brightest lensed image 
(in B 1933+507, B 2045+265, MG 0414+0534, PKS 1830-211 and other four 
unlabeled objects we have inferred $V$ from $I$ using $V-I=0.50$; 
in HST 1413+5211 and other unlabeled object we have inferred 
$V$ from $R$ using $V-R=0.11$). The data have been mainly obtained from the
CASTLES web page (http://cfa-www.harvard.edu/castles). For the previous
and subsequent calculations we have adopted from NED the visible
magnitude and redshift of NGC 5548, $m_V=13.3$ and $z=0.017$. To
transform apparent magnitudes into intrinsic luminosities we have made use
of the equation $L=4\pi D_{\rm lum}^{2} S (1+z)^{(\alpha-1)}$, which relates
the absolute luminosity of the source, the luminosity distance,
$D_{\rm lum}$, the apparent flux, $S$, and the spectral index, $\alpha$,
defined by $F_{\nu}\propto \nu^{-\alpha}$. We have considered an
$\Omega=0.3$ flat cosmology and a value for the spectral index
$\alpha=0.5$ (e.g., Richards et al. 2001). 

In Fig. \ref{fig1} we can identify a group of ten systems with magnitude
$m_V>21$ that would lie in the region $R \lesssim 100$ light-days 
(Balmer lines) and in the region $r \lesssim 10$
light-days (high-ionization lines). These systems are
possibly affected by microlensing, with particular strength in
the case of the high-ionization lines. However, the observed magnitudes
of the lensed QSOs should be corrected by extinction and
gravitational lens amplification. The amount of extinction in the
gravitational lenses is unknown for most of them, but it can be in the range from
dust-free lenses  to strongly reddened systems (e.g., Falco et al.
1999; Mu\~noz et al. 2001 and references therein). We selected the
magnitude from the brightest QSO image, which in most of the cases is
the less reddened. As a first approximation to the expected amount of
obscuration we can adopt the mean extinction, $\langle\Delta m\rangle 
=0.6$, derived
by Falco, Kochanek, \& Mu\~noz (1998) comparing the statistics of radio
and optical lensed quasars.  On the other hand, the quasar source is
amplified by the gravitational lens and the true luminosity of the
unlensed quasar should be calculated by fitting a lens model to each
system. The exact amplification of each gravitational lens system depends
on the lens model and on the particular configuration of the system, but
an average expected amplification  between the brightest image and the
unlensed source is $\sim 1.5$ mag (e.g., Leh\'ar et al. 2000).
Thus, a roughly averaged correction of the combined effects of extinction
and lens amplification can be made by adding $\sim 1$ mag to the
apparent QSO magnitude. 

Taking into account the +1 mag shift, the number of possible candidates 
affected by microlensing (at least in the high-ionization lines) is
increased to 12 ($\sim 40$\% of the sample, see Fig. 1). Our
selection procedure could include not only intrinsically low-luminosity but
also reddened objects in the list of candidates. As we have seen, a 1
mag shift due to underestimation of the extinction will not
substantially modify the set of candidates. This moderates the impact of
the extinction uncertainties in the selection of candidates, but we 
cannot discard the possibility
that our statistics were biased by heavily reddened objects
of intrinsically high luminosity. This is the case for B 0218+357, which
according to the rest-frame $E(B-V) = 1.52$ obtained from Falco et al.
(1999) would be an intrinsically high-luminosity object reddened by
extinction. MG 0414+0534 is also a very reddened object, but in this case
probably owing to a very red intrinsic spectral distribution ($F_{\nu}\propto
\nu^{-9}$ was measured at optical wavelengths by Hewitt et al. 1992). This 
implies that MG 0414+0534 could have a high intrinsic rest-frame V luminosity
but also indicates that this object is probably a radio galaxy rather than a QSO. 

The identification of the candidates available in the
literature supports the hypothesis
that most of them are intrinsically low-luminosity
objects, since in most cases the objects cannot be clearly classified as
bright QSOs but admit alternative classifications as objects with lower
levels of activity (radio galaxies, underluminous QSOs, starburst
galaxies, Seyfert 2 galaxies, or other types of AGNs). This also implies that we have probably
overestimated the luminosities of the candidate objects because we have
not removed the contribution from the galaxy, which is probably far from
negligible.

\subsection{Estimate of the amplification in B 1600+434}

It would be very useful to compute the expected 
amplifications for each object in the list of candidates. However, lack of knowledge 
of the extinction in each system and, to some extent, of a precise classification of 
the source, could affect the results significantly. We have selected the gravitational lens
discovered by Jackson et al. (1995), B 1600+434, a double-imaged typical quasar at $z_{\rm s}=1.59$, 
lensed by an edge-on spiral galaxy at $z_{\rm l}=0.42$,
in which both the lens amplification and the extinction might be reasonably well known. 
To calculate the amplification induced in B 1600+434 by the lens galaxy we have fitted
a singular isothermal ellipsoid (SIE) to this double-imaged quasar. Thus, we obtain an 
amplification of 0.96 mag for the brightest quasar 
image, which has an observed V-band magnitude of $m_V=22.69$. For the extinction 
we have used the result from Falco et al. (1999), $A_V=1.02$ mag. From the intrinsic luminosity
obtained 
after correcting for amplification and extinction ($m_V=22.63$) and using the Kaspi et
al. 
relationship, we have estimated radii of 45 and 5.3 ligth-days for the Balmer and
He~{\sc ii} BLRs, 
respectively (according to the two values of reference adopted for NGC 5548).

To estimate the maximum amplification, $\mu_{\rm max}$, we adopt 
the expression (e.g., Schneider et al. 1992):
\begin{equation}
\mu_{\rm max}\sim\frac{\sqrt{(\frac{r_{\rm BLR}}{\eta_0})^2+4}}{(\frac{ r_{\rm
BLR}}{\eta_0})},
\end{equation}
where $ r_{\rm BLR}$ is the BLR radius, and $\eta_0$ is given by
\begin{equation}
\eta_0=\sqrt{4\frac{GM}{c^2}\frac{D_{\rm s} D_{\rm ds}}{D_{\rm d}}}
\end{equation}
($D_{\rm d}$ and $D_{\rm s}$ are the angular diameter distances to the lens and the QSO 
respectively, and $D_{\rm ds}$ is the angular diameter lens--QSO distance).

If we apply these formulae to B 1600+434, we find that a solar-mass stellar
microlens would induce a modest, albeit potentially observable,
amplification of 1.4 in the Balmer lines and a strong amplification of
8.2 in He~{\sc ii}. A $0.1\ M_\odot$ microlens would also induce an appreciable
amplification of 2.76 in the He~{\sc ii} lines.  We do not calculate the
amplification by microlensing on the BEL for the other objects in the
list because the exact amount of extinction is unknown for most of them,
and because we can no longer suppose that the $F_{\nu}\propto
\nu^{-0.5}$ dependence of the energy distribution is a realistic approach
for some of the objects.

In summary, although most of the lensed quasars ($\gtrsim 70 \%$ of the
total) are intrinsically high-luminosity quasars (so that no strong
microlensing in the BEL is expected) we found that $\lesssim 30\%$ of
the lensed sources are faint enough to be considered as possibly
affected by microlensing in the BEL.

\section{MICROLENSING EFFECTS ON THE PROFILE OF THE BROAD
EMISSION LINES}
 
When an organized velocity field governs the kinematics of the BLR, microlensing can
give rise to a selective amplification of the emission-line profile (Nemiroff 1988). The shape of
the line depends on the location of the source with respect to the optical axis (defined by the
observer and the microlens) and can change with the relative movement between the
microlens and the BLR. In this section we adopt the kinematic models for the BLR used by
Robinson (1995) to study the range of profile shapes that exists among AGN in the context of a
simple parameterization of some basic properties of the BLR such as emissivity and velocity
law. Our intention is to study, in this framework, the effects induced in the line profile by a
microlens in different locations with respect to the center of the BLR. 

We  assume that a single microdeflector is affecting  the BLR. This supposition  
adequately serves the objectives of the present paper (a more realistic approach based
on the existence of a random distribution of microdeflectors will be analyzed in a future paper).

In accordance with Robinson (1995), we consider three different geometries (spherical,
biconical, and cylindrical) and adopt the following radial dependences for the emissivity and  
magnitude of the velocity:
\begin{equation}
\label{ecemis}
\epsilon(r)=\epsilon_0 \left (\frac{r}{r_{\rm in}}\right )^{\beta}
\end{equation}
and
\begin{equation}
v(r)=v_0\left (\frac{r}{r_{\rm in}}\right )^{p}.
\end{equation}

We compute the emission-line profile from the expression
\begin{equation}
F _\lambda=\int_V \epsilon(r) ~ \delta \left [\lambda-\lambda_0
\left (1+\frac{v_{\shortparallel}}{c}\right )\right ] ~ \mu(\vec{r}) ~ dV ,
\end{equation}
where 
\begin{equation}
\label{ecampli}
\mu(\vec{r})=\frac{u^2+2}{u\sqrt{u^2+4}} \quad \left(
u=\frac{|\vec{r}-\vec{r}_0|}{\eta_0}\right)
\end{equation}
is the amplification associated with the microlens, $\vec{r}_0\sim(r_0,\varphi_0)$ is the
position of the microlens, $\eta_0$ is the Einstein radius, and $v_{\shortparallel}$ is the projected 
line-of-sight velocity.

We  adopt inner ($r_{\rm in}$) and outer ($r_{\rm BLR}$) radii for the BLR.

It is convenient to scale the distances to the Einstein radius associated 
with the microlens.
In this way, the relevance of the microlens effect can be characterized by the quantity 
\begin{equation}
\tilde{r}_{\rm BLR}=\frac{r_{\rm BLR}}{ \eta_0}.
\end{equation}

We consider two values (1 and 4) for $\tilde{r}_{\rm BLR}$. We assume throughout that
$r_{\rm in}=0.1\ r_{\rm BLR}$.

To study the effects produced by the relative off-centering between the microlens and
the BLR, we  compute line profiles corresponding to a grid of displacements of the
microlens relative to the center of the BLR. We consider 25 positions in the positive 
$XY$ quadrant ranging from $0$ to $\tilde{r}_{\rm BLR}$ in both the $X$ and $Y$
axes (Fig. \ref{fig2}). 

In the Appendix we collect all the formal development and formulae and   concentrate 
on the results in the following sections.

\subsection{SPHERICAL SHELL} 

In the case of a spherical ensemble of emitters flowing radially, the projected 
line-of-sight velocity corresponding to a emitter at a position $(r,\theta)$ is given by
\begin{equation}
v_{\shortparallel}=v_0 \left (\frac{r}{r_{\rm in}}\right )^p \cos \theta\quad
\left(p>0\right).
\end{equation}

We have used the same notation as Robinson (1995) for the parameters related to the 
relative velocity,
\begin{equation}
x=\frac{\lambda-\lambda_0}{\lambda_0}\frac{c}{v_{\rm max}},
\end{equation}
and we will refer to the line parameter which defines the line shape:
\begin{equation}
\label{eceta}
\eta=\frac{\beta+3}{p}-1.
\end{equation}

The line profile (see Appendix A.1) is obtained by integrating
\begin{equation}
\label{ecflux}
\ F_x  = \left\{
  \begin{array}{ll}
        \frac{\epsilon_0 r_{\rm int}^2 c}{\lambda_0 v_0} 
        \int_0^{2\pi} \int_{{\rm Max}[r_{\rm in}, r_{\rm lim}]}^{r_{\rm BLR}} 
        \left(\frac{r}{r_{\rm in}}\right)^{\beta +2 -p} [\mu(x,r,\varphi)]_{f=0}  dr d\varphi
& \mbox {$r_{\rm lim}<r_{\rm BLR}$ } \\
      0 & \mbox {$r_{\rm lim}>r_{\rm BLR}$ }  
    \end{array}
  \right.        
\end{equation}
where  $r_{\rm lim}= r_{\rm in}~ (x/{x_{\rm m}})^{1/p}$, $x_{\rm m}=v_0/v_{\rm
max}=(r_{\rm in}/r_{\rm
BLR})^p$, and 
$[\mu(x,r,\varphi)]_{f=0}$ is given by Eq. \ref{ecampli}, inserting 
Eq. \ref{apsins} into Eq. \ref{apesfb}.

Following Robinson (1995), we  take $\eta$ as the parameter defining the
emission-line profile. 
For the spherical case we consider two values, a) $\eta=2$ and 
b) $\eta=-0.25$, which in absence of microlensing would correspond to concave and
convex profile wings, respectively. In absence of microlensing, only this parameter is needed 
to characterize the line profile. In the presence of microlensing we need also to fix
another parameter. We select the emissivity exponent $\beta=-1.5$ (see Eq. \ref{ecemis}). The
exponent of the velocity law, $p$, is then obtained from Equation \ref{eceta}. Thus, for case 
a) we have $p=0.5$ and for case b) $p=2$.

In Figs. \ref{fig3}, \ref{fig4} we present the grids of profiles for $\tilde{r}_{\rm BLR}=1,4$ for
case a) and in Figs. \ref{fig5}, \ref{fig6} for case b). The influence of microlensing on the line profiles would 
be observable when the microlens is centered on the BLR, and also in 
many other cases in which the microlens is off-centered. The displacement of a
microlens across the BLR would induce changes in the relative strength of different parts of the 
line profile, relatively enhancing the wings when the microlens is centered on the BLR 
and the core when it is sited in the outer parts.  However, no relative enhancements 
between the blue and red parts of the profile would appear due to the high symmetry of 
the spherical model.

As mentioned previously, one notable property of Robinson's models is that the 
profile shape  depends only on $\eta$. This implies that from the study of the line profile
it is not possible to derive direct information on the velocity field or the emissivity
law. Could the presence of microlensing break this degeneracy for a spherical shell?
Such questions as the existence of a monotonically increasing or decreasing dependence with
radial distance of the velocity field may have a formal answer. In principle, it would be
possible to invert the line-profile expression (Eq. \ref{ecflux}) obtained for several different
positions of the microlens to recuperate, making suitable suppositions, the law for radial velocities.  
However, it is not easy to decide on direct observational criteria to carry out this study.

To illustrate the difficulties in deriving information from the microlensed line profile in 
the spherical case, notice that not even a simple question like the existence of outflow
or inflow can be answered. (For the spherical case there is always the same contribution 
of receding and approaching emitters along the line of sight which would undergo the
same magnification.)

\subsection{BICONICAL SHELL}

Much observational evidence (Zheng, Binette, \& Sulentic 1990; Marziani, Calvani, \& Sulentic 1992)
and theoretical work supports the idea that the flow of emitting gas is anisotropic, preferentially 
confined to a pair of oppositely directed cones. In this model we need three polar coordinates 
$(r,\theta,\varphi)$ measured with respect to the cone axis to express the projected
line-of-sight velocity corresponding to an emitter,
\begin{mathletters}
\begin{equation}
v_{\shortparallel}=v_0 \left (\frac{r}{r_{\rm in}}\right )^p \xi\quad (p>0),
\end{equation}
\begin{equation}
\xi=\sin \theta \sin \varphi \sin i + \cos \theta \cos i,
\end{equation}
\end{mathletters}
where $i$ is the inclination of the cones axis with respect to the line of sight.

This model can give rise to a variety of line profiles (see Figure 5 of Robinson 1995). 
We are going to consider the two limiting cases: $i=0^\circ$ and $i=90^\circ$. We will 
adopt the cone half-angle $\theta_{\rm c}=30^\circ$, and $\beta=-1.5$.

\subsubsection{$i=0^\circ$}

The line profile (see Appendix A.2.1) is obtained by integrating
\begin{equation}
\ F_x  = \left\{
  \begin{array}{ll}
        \frac{\epsilon_0 r_{\rm int}^2 c}{\lambda_0 v_0} 
        \int_0^{2\pi} \int_{{\rm Max}[r_{\rm lim},r_{\rm in}]}^{{\rm Min}[r_{\rm
sup},r_{\rm BLR}]}
        \left(\frac{r}{r_{\rm in}}\right)^{\beta +2 -p} [\mu(x,r,\varphi)]_{f=0} dr d\varphi  
        & \left(\mbox {$r_{\rm lim}<r_{\rm BLR} ~{\rm and}~r_{\rm sup}>r_{\rm
in}$}\right) \\
      0 & \mbox {in the other cases }  
    \end{array}
  \right.        
\end{equation}
where $r_{\rm lim}= r_{\rm in}~ (|x|/{x_{\rm m}})^{1/p}$, $r_{\rm sup}=r_{\rm
BLR}~(|x|/\cos\theta_{\rm c}
)^{1/p}$,
and $[\mu(x,r,\varphi)]_{f=0}$ is given by Eq. \ref{ecampli}, inserting 
Eq. \ref{apsins} into Eq. \ref{apesfb},
with $x_{\rm m}=v_0/v_{\rm max}=(r_{\rm in}/r_{\rm BLR})^p$.

In Figs. \ref{fig7}, \ref{fig8} we present the grids of profiles for $\tilde{r}_{\rm BLR}=1,4$
corresponding to $\eta=2$, and in Figs. \ref{fig9}, \ref{fig10} the grids corresponding to $\eta=-0.25$. 
In both
cases we obtain two-peaked profiles. As in the case of the sphere, the influence of
microlensing when the microdeflector is centered with the BLR is very noticeable. However, in the
biconical case, the change in the relative enhancements of 
different parts of the line profile caused by the displacement of the microlens with
respect to the center of the BLR is more noticeable than in the spherical one.

\subsubsection{$i=90^\circ$}

The line profile (see Appendix A.2.2) is obtained by integrating
\begin{equation}
F_x  =\int\limits_{{\rm Max}[r_{\rm lim},r_{\rm in}]}^{r_{\rm BLR}}\left (\left [
        \int\limits_{\theta_{\rm lim}}^{\theta_{\rm c}} +
        \int\limits_{\pi-\theta_{\rm c}}^{{\rm Min}[\pi-\theta_{\rm lim},\pi]}~ \right ]
         f(x,r,\theta) ~d\theta \right )dr,
\end{equation}
where
\begin{equation}
\ f(x,r,\theta)  = \left\{
  \begin{array}{ll}
        \frac{\epsilon_0 r_{\rm in}^2 c}{\lambda_0 v_0} ~\left( \frac{r}{r_{\rm
in}}\right)^{\beta +2 -p}
          \frac{\sin \theta ~[\mu_+(x,r,\theta)+\mu_- (x,r,\theta)]_{f=0}}{\sqrt{\sin ^2
\theta 
          -\left[ \frac{x}{x_{\rm m}} \left(\frac{r}{r_{\rm in}}\right)^{-p}\right]^2}}
        &\left(  \mbox {$\theta>\theta_{\rm lim}$}\right), \\
      0 & \mbox {in other cases},  
    \end{array}
  \right.        
\end{equation}
with $\theta_{\rm lim}=\arcsin[ (|x|/x_{\rm m}) (r/r_{\rm in})^{-p}]$,
$x_{\rm m}=v_0/v_{\rm max}=(r_{\rm in}/r_{\rm BLR})^p$, and 
$[\mu_{\pm}(x,r,\theta)]_{f=0}$ is given by Eq. \ref{ecampli}, inserting Eq. \ref{apcosb} into
Eq. \ref{apmodc}.

In Figs. \ref{fig11}, \ref{fig12} we present the grids of profiles for $\tilde{r}_{\rm BLR}=1,4$
corresponding to $\eta=2$ and in Figs. \ref{fig13}, \ref{fig14} the grids corresponding to $\eta=-0.25$. In this
case, $i=90^\circ$, we obtain single-peaked profiles, convex for $\eta=2$, and concave for 
$\eta=-0.25$. The effects of microlensing are very strong, 
even for the case $\eta_0=r_{\rm BLR}/4$ (Figs. \ref{fig12} and \ref{fig14}).

For the limiting geometries  considered here ($i=0^\circ, 90^\circ$) there are no
asymmetries induced by the microlensing in the line profile. However, for an arbitrary inclination, 
an off-centered microlens should induce relative enhancements of the blue and red
parts. Due to the loss of symmetry of the projected velocity field with respect to that of the spherical 
case, the inversion of the profile equation to study the velocity field should be easier. For
instance, in the case of small cone aperture, the crossing of a microlens along the
biconical axis would serve to virtually map the radial dependence of the velocity field. 

\subsection{CYLINDRICAL SHELL}

Rotation in a plane disk has been often considered in relation to the kinematics of the 
BLR and is typically characterized by the presence of a central dip (Mathews 1982) in
the line profile arising from the finite extension of the outer radius of the disk. This feature is 
not usually found in the observed line profiles, but there are several ways to avoid it in 
the models (see Robinson 1995).  Nevertheless, the existence of rotation is strongly
supported by recent observational work (Peterson \& Wandel 1999). 

\subsubsection{Keplerian disk}

The Keplerian disk is a particular case of the cylindrical disk with velocity field,
\begin{equation}
v_{\shortparallel}=v_0 \left (\frac{r}{r_{\rm in}}\right )^p \cos \varphi \sin i\quad
(p=-0.5),
\end{equation}
where $r$ and $\varphi$ are polar coordinates of an emitter in the disk. 

The line profile is given by (see Appendix A.3.1):
\begin{equation}
\ F_x  = \left\{
  \begin{array}{ll}
        \frac{\epsilon_0 r_{\rm in} c}{\lambda_0 v_0 \sin i } 
        \int_{r_{\rm in}}^{{\rm Min}[r_{\rm lim},r_{\rm BLR}]} \left( \frac{r}{r_{\rm
in}}\right)^{\beta+1-p}\frac{
        ~[\mu_+(x,r) +\mu_-(x,r) ]_{f=0}}{\sqrt{1- \left [\frac{x}{\sin
i}\left(\frac{r}{r_{\rm in}}\right)^{-p}
        \right] ^2}} ~dr
        & \mbox {$(r_{\rm lim} >r_{\rm in})$} \\
        0      
        & \mbox {$(r_{\rm lim} <r_{\rm in})$}  
    \end{array}
  \right.        
\end{equation}
where $r_{\rm lim}= r_{\rm in}~ (|x|/\sin i)^{1/p}$, $x_{\rm m}=(r_{\rm in}/r_{\rm
BLR})^{-p}$ and 
 $[\mu_{\pm} (x,r)]_{f=0}$ is given by Eq. \ref{ecampli}, inserting Eqs. \ref{apcosc} and
\ref{apsinc} into Eq. \ref{apmodd}.

In Figs. \ref{fig15}, \ref{fig16} we present the grid of profiles for $\tilde{r}_{\rm BLR}=1,4$ with 
$i=45^{\circ}$, $\beta=-1.5$, and $\eta=(\beta +2)/p - 1$. 
The most noticeable feature associated with the Keplerian case (and with the 
cylindrical case  in general) is the presence of strong asymmetries in the line profiles
induced by the microlensing. 

\subsubsection{Modified Keplerian disk}

An easy way of generating a single-peaked profile in the cylindrical case is to modify 
the velocity field increasing the contribution of low velocities. For our purpose we will adopt
\begin{mathletters}
\begin{equation}
v_{\shortparallel} = v(r) \cos\varphi \sin i\quad (p=-0.5)
\end{equation}
and
\begin{equation}
v(r) = v_0  \left (\frac{\frac{1}{r}-\frac{1}{r_{\rm BLR}}} {\frac{1}{r_{\rm
in}}-\frac{1}{r_{\rm BLR}}}
\right )^{-p} =v_0 ~u(r),
\end{equation}
\end{mathletters}
where $r$ and $\varphi$ are polar coordinates of an emitter in the disk. 

In this case, the line profile is given by (see Appendix A.3.2),
\begin{equation}
\ F_x  = \left\{
  \begin{array}{ll}
        \frac{\epsilon_0 r_{\rm in} c}{\lambda_0 v_0 \sin i } 
        \int_{r_{\rm in}}^{{\rm Min}[r_{\rm lim},r_{\rm BLR}]} \left( \frac{r}{r_{\rm
in}}\right)^{\beta+1}\frac{
        ~[\mu_+(x,r) +\mu_-(x,r) ]_{f=0}}{u(r)\sqrt{1- \left (\frac{x}{u(r)\sin i}
        \right) ^2}} dr
        & \mbox {$\left ( r_{\rm lim} >r_{\rm in}\right )$} \\
        0      
        & \mbox {$\left( r_{\rm lim} <r_{\rm in}\right) $}  
    \end{array}
  \right.        
\end{equation}
where $r_{\rm lim}=\left[ \left(|x|/\sin i\right)^{-1/p}\left(1/r_{\rm in}-1/r_{\rm
BLR}\right)
+1/r_{\rm BLR}\right]^{-1}$, $x_{\rm m}=r_{\rm in}/r_{\rm BLR}$, and
$[\mu_{\pm} (x,r)]_{{f=0}}$ is given
by Eq. \ref{ecampli}, inserting Eqs. \ref{apcosc} and
\ref{apsinc} into Eq. \ref{apmodd}.

In Figs. \ref{fig17}, \ref{fig18} we present the grid of profiles for $\tilde{r}_{\rm BLR}=1,4$ with
$i=45^{\circ}$and $\beta=-1.5$. In addition to the asymmetries, the most interesting effect 
of microlensing on the line profiles corresponding to the Keplerian modified velocity field
is the displacement of the line peak with respect to the centroid of the non-microlensed line
profile, $\Delta x_{\rm p}$, (Fig. \ref{fig19}), which, 
independently of the mass considered ($\tilde{r}_{\rm BLR}=1,4$), can be of the order of as
much as $\sim \rm FWZI/4$.

\section{DISCUSSION}

If the models and assumptions of the previous sections constitute a good
description  of  BLR microlensing, we could conclude that the effects of
this phenomenon on the line  profiles should be not only noticeable but
also easily detectable in some lens systems. 

The experimental situation, however, is more complex. In first place, the BELs
are blended  with the narrow emission lines, which come from the much more extended
narrow-line region (NLR),  which would be not affected by microlensing. 
In fact, it is likely that diverse transition  regions between the BLR and the
NLR could also contribute to the core of the line profile.  Second, the
compactness of the lens systems makes  observation very difficult.  For
instance, in the case (B~1600+434) 
in which we have obtained a realistic estimate for microlensing amplification  there are
two compact images, the lens galaxy (an edge-on spiral), and some additional 
extended emission, all within a separation
of $\sim 2''$.  This  is a major setback when trying to obtain
individual spectra, and only modern spectroscopic  techniques (2D spectroscopy,
Mediavilla {et al.} 1998 and references therein) or observation from space
avoid the problems induced by source blending and  differential atmospheric
refraction. In any case, to detect microlensing we should obtain spectra
(preferably of high-ionization lines) with a high S/N ratio in the wings of the
line profile, where the contribution from the BLR would be  dominant.

The emissivity is another important parameter that can affect the detection of 
microlensing in BELs. In Fig. \ref{fig20} we present the ratio of the amplified to
unamplified line profiles corresponding to the modified Keplerian 
case, where we have changed the emissivity parameter, $\beta$. For a highly
concentrated BLR ($\beta =-4$) the effect is stronger when the BLR is almost centered 
on the microlens and weak for larger displacements of the microlens with respect to 
the center of the BLR. But when the emissivity is constant (for a disk of uniform
brightness, $\beta=0$) the effects of microlensing remain noticeable for all 
displacements considered and are hence more likely to be observed.

In Fig. \ref{fig1} we have labeled the gravitational lens systems in which
significant BEL microlensing could be  detected (30\% of the total).
However, in other systems a modest effect could be detectable by looking at
high-ionization lines. This is important because other questions than the
amplification, such as the crossing time of the microlens across the BLR or
the frequency of the events, can lead us to study a gravitational lens
system in particular. For instance, in a favorable case it would be
possible not only to compare the line profile  corresponding to
microlensed and non-microlensed images but also to observe in the line 
profile of an image changes attributable to the microlens crossing. In
the most favorable case from this perspective, Q 2237+0305, microlensing
events are reported each year with crossing times of the order of a year
or less. However, this is a bright QSO in which only very modest
amplifications of 30\% or less can be expected in the high-ionization
lines. This estimation of the amplification is, in any case, subject to changes in the BLR-size vs
QSO-luminosity relationship, and also to the expected intrinsic
dispersion from object to object.

In spite of the last comment, the amplifications associated with the
high-ionization lines could be high in many cases, and only a strong
departure from the assumptions made in this study (e.g., a severe
underestimate of the extinction, very different behavior of the
BLR-size/QSO-magnitude relationship, or an unexpectedly low-mass
population of microlenses) could avoid the detection of microlensing in
the BELs by comparing the high-ionization line profiles of a microlensed
and a non-microlensed image in the most favorable cases. According to
this, the study of the incidence of microlensing in BELs could become a
tool for studying the BLR size and stratification, especially when lines of
different ionization are observed.

\section{CONCLUSIONS}

In the light of recent discoveries concerning the BLR size and its scaling with AGN luminosity, 
currently accepted values for  microlens masses, and a variety of kinematic models for the BLR, 
we have revisited the influence of microlensing in the BEL. We have computed grids of line 
profiles corresponding to different displacements of the microlens with respect to the BLR center. 
Some results are worth  summarizing:

1.- The global amplification of the BEL induced by microlensing events
could be relevant. We identify a group of ten gravitational lens
systems (about 30\% of the total sample) for which the microlensing
effect could be observable, especially in high-ionization lines. In
other gravitational lenses the microlensing amplifications would be much
more modest.

2.- Even for relatively small microlenses corresponding to high values
of the  BLR radius/Einstein radius quotient, ($r_{\rm BLR}/\eta_0\sim
4$), the effects produced by the differential  amplification of the line
profile (relative enhancement of different parts of the line profile, 
line asymmetries, displacement of the peak of the line, etc.) would be
easily detectable except  for highly symmetric velocity fields. The
displacement of the peak of the line profile caused by  microlensing is
especially interesting, since it could otherwise induce  inexplicable
redshift  differences between the different images in a gravitational
lens system. 

3.- The study of the changes between the BEL profiles corresponding to microlensed
and non-microlensed images, or among the BEL profiles of lines with different ionization in
a microlensed image could be useful for probing current ideas about BLR size and stratification.

\acknowledgments

We would like to acknowledge valuable and useful comments
of the anonymous referee. This work was supported by the P6/88    
project of the Instituto de Astrof\'\i sica de Canarias (IAC).
L. \v C. P. is supported by Ministry of Science, Technologies and
Development of Serbia through the project `Spectroscopy of Extragalactic
Objects'.

\clearpage

\appendix

\section{MODELING THE BROAD LINE REGION}

Nowadays, there is no generally accepted model for the geometry 
and dynamics of the broad line region. So, we adopt in this 
paper the discrete cloud model (see Robinson 1995) in which it is 
assumed that the magnitude of the velocity and the volume 
emissivity may be independently specified by a power law. It is 
also assumed that the clouds or volume elements emit line 
radiation isotropically.

The amplified line profile produced by the system is obtained 
from:
\begin{equation}
\label{apflux}
F_{\lambda} = \int\limits_V \epsilon(r)~ \delta \left [ 
\lambda - \lambda_0 \left ( 1 + \frac{v_{\shortparallel}}{c} 
\right )\right ] \mu (\vec{r}) dV ,
\end{equation}
where the integral is taken over the total volume occupied by 
the clouds, $\mu (\vec{r})$ is the relative amplification that 
the microlens produces over the line profile, and 
$v_{\shortparallel}$ is the projection of the velocity field 
along the line of the sight. Finally, $\epsilon(r)$ is 
the volume emissivity of the clouds:
\begin{equation}
\epsilon(r) = \epsilon_0 \left( \frac{r}{r_{\rm in}} 
\right )^{\beta},
\end{equation}
$r_{\rm in}$ being the inner radius of the cloud system.

The geometry of the BLR is unknown, but it 
seems plausible that its overall structure could be 
characterized by one of three basic symmetries: spherical, 
conical, or cylindrical.

\subsection{SPHERICAL SHELL}

Let us first consider a spherically symmetric cloud ensemble of 
outer radius $r_{\rm BLR}$, which is characterized by a radial 
velocity field. The component of the velocity parallel to the 
line of sight is:
\begin{equation}
v_{\shortparallel} = v_0 \left ( \frac{r}{r_{\rm in}} \right )^p 
\cos\theta\quad (p>0).
\end{equation}

If the microlens is placed at $\vec{r}_0\sim(r_0,\varphi_0)$ and 
the Einstein radius is $\eta_0$, the relative amplification is 
given by (e.g., Schneider et al. 1992):
\begin{mathletters}
\label{apampli}
\begin{equation}
\label{apamplia}
\mu(\vec{r}) = \frac{u^2+2}{u\sqrt{u^2+4}},\quad\quad
u = \frac{|\vec{r}-\vec{r}_0|}{\eta_0} ~ ,
\end{equation}
\begin{equation}
\label{apamplib}
|\vec{r} - \vec{r}_0| = \sqrt{(r \sin\theta \cos\varphi - 
r_0 \cos\varphi_0)^2 + (r \sin\theta \sin\varphi - r_0 
\sin\varphi_0)^2}.
\end{equation}
\end{mathletters}

To compute the integral in Eq. \ref{apflux}, we define:
\begin{equation}
f \equiv \lambda - \lambda_0 \left [ 1 + \frac{v_0}{c}\left ( 
\frac{r}{r_{\rm in}} \right )^p \cos\theta \right ] .
\end{equation}

Equation \ref{apflux}, when the integral in the $\theta$ dimension is 
done and the parameters $x=c~(\lambda-\lambda_0)/( v_{\rm max}\lambda_0)$ 
and $x_{\rm m}=v_0/v_{\rm max}=(r_{\rm in}/r_{\rm BLR})^p$ are 
defined, appears as:
\begin{equation}
\ F_x  = \left\{
  \begin{array}{ll}
        \int_{0}^{2 \pi} \int_{{\rm Max}[r_{\rm lim},r_{\rm in}]}^{r_{\rm BLR}} 
        \epsilon(r) r^2 \frac{[\mu (x,\vec{r})]_{f=0}}{[\frac{\rm df}{d\theta}]_{f=0}}
        [\sin\theta]_{f=0} drd\varphi  & \mbox {$\left(r_{\rm lim}<r_{\rm BLR}\right)$ }
\\
      0 & \mbox {$\left(r_{\rm lim}>r_{\rm BLR}\right)$} 
    \end{array}
  \right.        
\end{equation}
with
\begin{equation}
[\cos\theta]_{f=0} = \frac{x}{x_{\rm m}} \left ( \frac{r}{r_{\rm in}}
\right )^{-p},
\end{equation}
\begin{equation}
\label{apsins}
[\sin \theta]_{f=0} = + \sqrt{1-[\cos\theta]_{f=0}^2},
\end{equation}
and
\begin{equation}
\label{apdfs}
\left [ \frac{\rm df}{d\theta}\right ]_{f=0} = \lambda_0 
\frac{v_0}{c} \left ( \frac{r}{r_{\rm in}} \right )^p 
[\sin\theta]_{f=0},
\end{equation}
where $[\mu(x,\vec{r})]_{f=0}$ is given by Eq. \ref{apamplia},
inserting Eq. \ref{apsins} into Eq. \ref{apamplib} as:
\begin{equation}
\label{apesfb}
|\vec{r} - \vec{r}_0|_{\pm, f=0} = \sqrt{(r [\sin \theta]_{f=0} \cos\varphi - 
r_0 \cos\varphi_0)^2 + (r [\sin \theta]_{f=0} \sin\varphi - r_0 
\sin\varphi_0)^2},
\end{equation}
and $r_{\rm lim}= r_{\rm in}~ (|x|/{x_{m}})^{1/p}$. The expression of 
$r_{\rm lim}$ is obtained from the delta condition.

\subsection{BICONICAL SHELL}

In this case, the broad lines are emitted by a bipolar flow in 
which clouds moving radially outwards are confined to a pair of 
oppositely directed cones, whose common axis passes through the 
optical axis. In general, an individual cloud measured with 
respect to the cone axis has a projected velocity:
\begin{mathletters}
\begin{equation}
v_{\shortparallel} = v_0 \left ( \frac{r}{r_{\rm in}} \right )^p \xi
\quad (p>0),
\end{equation}
\begin{equation} 
\xi = \sin\theta \sin\varphi \sin i + \cos\theta \cos i,
\end{equation}
\end{mathletters}
where $i$ represents the inclination of the axis with respect to 
the line of sight. For this system, the amplified line profile is:
\begin{equation}
\label{apfluxb}
F_{\lambda} = \int_{0}^{2\pi} \int_{r_1}^{r_2}
\left [\int_{0}^{\theta_{\rm c}}+\int_{\pi -\theta_{\rm c}}^{\pi}\right ] \epsilon(r) ~
\delta \left [ \lambda - \lambda_0 \left ( 1 
+ \frac{v_{\shortparallel}}{c}  \right )
\right ] \mu (\vec{r}) r^2 \sin\theta d\theta dr d\varphi,
\end{equation}
where the amplification is given by Eq. \ref{apamplia}, with:
\begin{equation}
\label{apmodb}
|\vec{r}-\vec{r}_0| = \sqrt{( r \sin\theta \cos\varphi - r_0 
\cos\varphi_0 )^2 + ( r \sin \theta \sin \varphi \cos i - r \cos \theta \sin i- r_0
\sin\varphi_0 )^2}.
\end{equation}

\subsubsection{$i=0^\circ$ CASE}

The $i=0^\circ$ case leads to the amplified line profile:
\begin{equation}
\ F_x  = \left\{
  \begin{array}{ll}
         \int_0^{2\pi}\int_{{\rm Max}[r_{\rm lim},r_{\rm in}]}^{{\rm Min}[r_{\rm
sup},r_{\rm BLR}]}
         \epsilon(r) r^2  \frac{[\mu (x,\vec{r})]_{f=0}}{[\frac{\rm df}{d\theta}]_{f=0}}
        [\sin\theta]_{f=0} dr d\varphi
        & \mbox {$\left( r_{\rm lim}<r_{\rm BLR}\ {\rm and}\ r_{\rm sup}>r_{\rm
in}\right)$}, \\
      0 & \mbox {in the other cases},  
    \end{array}
  \right.        
\end{equation}
where $[\sin\theta]_{f=0}$ and $[df/d\theta]_{f=0}$ are given by Eqs. \ref{apsins} and
\ref{apdfs}, 
$[\mu(x,\vec{r})]_{f=0}$ is given by  Eq. \ref{apamplia}, inserting Eq. \ref{apsins} into Eq.
\ref{apesfb}, with
$r_{\rm lim}= r_{\rm in}~ (|x|/{x_{\rm m}})^{1/p}$, and $r_{\rm sup}=r_{\rm
BLR}~(|x|/\cos\theta_{\rm c}
)^{1/p}$. 
The expressions of $r_{\rm lim}$ and $r_{\rm sup}$ can be inferred from delta
conditions.

\subsubsection{$i=90^\circ$ CASE}

In the $i=90^\circ$ case we consider a projected velocity of
\begin{equation}
v_{\shortparallel} = v_0 \left ( \frac{r}{r_{\rm in}} \right )^p 
\sin\theta \sin\varphi \quad (p>0).
\end{equation}

Let us now define $f$ as:
\begin{equation}
f \equiv \lambda - \lambda_0 \left [ 1 + \frac{v_0}{c} \left (
\frac{r}{r_{\rm in}} \right )^p \sin\theta \sin\varphi \right ],
\end{equation}
and then, after integrating in the $\varphi$ dimension and 
adopting that $x=c~(\lambda-\lambda_0)/( v_{\rm max}\lambda_0)$ 
and $x_{\rm m}=v_0/v_{max}=(r_{\rm in}/r_{\rm BLR})^p$, Eq. \ref{apfluxb}
becomes:
\begin{equation}
\ F_x  =
        \int\limits_{{\rm Max}[r_{\rm lim},r_{\rm in}]}^{r_{\rm BLR}}\left (\left [
        \int\limits_{\theta_{\rm lim}}^{\theta_{\rm c}} +
        \int\limits_{\pi-\theta_{\rm c}}^{{\rm Min}[\pi -\theta_{\rm lim},\pi]}\right ]
         f(x,r,\theta)  d\theta \right )dr,
\end{equation}
where
\begin{equation}
\ f(x,r,\theta)  = \left\{
  \begin{array}{ll}
        \epsilon(r)  r^2 \sin\theta \frac{[\mu_+ 
        (x,\vec{r})+\mu_- (x,\vec{r})]_{f=0}}{[\frac{\rm df}{d\varphi}]_{-, f=0}},  
        & \mbox {$\left(\theta>\theta_{\rm lim}\right)$} \\
      0 & \mbox {in other cases}  
    \end{array}
  \right.        
\end{equation}
with $\theta_{\rm lim}=\arcsin[ (|x|/x_{\rm m}) (r/r_{\rm in})^{-p}]$.
The expression of $\theta_{\rm lim}$ is inferred from delta conditions.
Moreover,
\begin{equation}
[\sin \varphi]_{f=0} = \frac{x}{x_{\rm m}} \left ( \frac{r}{r_{\rm in}}
\right )^{-p} \frac{1}{\sin\theta},
\end{equation}
\begin{equation}
\label{apcosb}
[\cos \varphi]_{\pm,f=0} = \pm \sqrt{1-[\sin\varphi]_{f=0}^2},
\end{equation}
and
\begin{equation}
 \left [\frac{\rm df}{d\varphi}\right ]_{\pm, f=0}  = 
     -\lambda_0  \frac{v_0}{c} \left ( \frac{r}{r_{\rm in}} \right )^p \sin\theta ~
     [\cos \varphi]_{\pm,f=0}.
\end{equation}
$[\mu_{\pm}(x,\vec{r})]_{f=0}$ is given by Eq. \ref{apamplia}, inserting Eq. \ref{apcosb} into
Eq. \ref{apmodb} as
\begin{equation}
\label{apmodc}
\ |\vec{r}-\vec{r}_0|_{\pm, f=0}  = 
  \sqrt{(r \sin\theta [\cos \varphi]_{\pm,f=0} - r_0 \cos\varphi_0)^2+(r\cos \theta+r_0\
sin\varphi_0)^2}.      
\end{equation}

\subsection{CYLINDRICAL SHELL}

The simplest example of cylindrical symmetry is a plane, i.e., a 
thin disk. We will suppose that this disk has  uniform 
thickness, $h\ll r_{\rm in}$, and that the angle between its axis and 
the line of sight is $i$. Finally, any point in the disk is assumed 
to follow a circular orbit about the axis, thus giving a 
line-of-sight velocity of:
\begin{equation}
v_{\shortparallel} = v(r) ~ \cos\varphi \sin i,
\end{equation}
where $r$ and $\varphi$ are polar coordinates of an emitter in the disk. 

For this system, the amplified line profile is
\begin{equation}
\label{apfluxc}
F_{\lambda}= \int_S 
\epsilon (r)~ \delta \left [\lambda - \lambda_0 \left ( 1+\frac{v_{\shortparallel}}
{c}\right )\right ] \mu (\vec{r})~ dS.
\end{equation}

Let us now define $f$ as:
\begin{equation}
f \equiv \lambda - \lambda_0 \left [ 1 + \frac{v(r)\cos\varphi \sin i }{c} 
\right ],
\end{equation}
and $x=c~(\lambda-\lambda_0)/( v_{\rm max}\lambda_0)$, with $v_0=v_{\rm max}$.
Then, when the integral in the $\varphi$ dimension is done, Eq. \ref{apfluxc} becomes:
\begin{equation}
\ F_{x}  = \left\{
  \begin{array}{ll}
      \int_{r_{\rm in}}^{{\rm Min}[r_{\rm lim},r_{\rm BLR}]} \epsilon (r) r\frac{[\mu_+
(x,\vec{r})+
      \mu_- (x,\vec{r})]_{f=0}} {[\frac{df}{d\varphi}]_{+,f=0}} dr      
      & \mbox {$\left( r_{\rm lim}>r_{\rm in}\right)$}\\
      0
      & \mbox {$\left ( r_{\rm lim}<r_{\rm in}\right)$}  
    \end{array}
  \right.        
\end{equation}
where
\begin{equation}
\label{apcosc}
[\cos\varphi]_{f=0} = x\frac{v_0 }{v(r) \sin i},
\end{equation}
\begin{equation}
\label{apsinc}
[\sin\varphi]_{\pm,f=0} = \pm \sqrt{1-[\cos\varphi]_{f=0}^2} ,
\end{equation}
and
\begin{equation}
 \left [\frac{df}{d\varphi}\right ]_{\pm, f=0}  = 
     \lambda_0 \frac{v(r) \sin i}{c} ~[\sin\varphi]_{\pm,f=0}.
\end{equation}
$[\mu_{\pm}(x,\vec{r})]_{f=0}$ is given by Eq. \ref{apamplia}, inserting the 
expressions of $[\cos\varphi]_{f=0}$ and $[\sin\varphi]_{f=0}$ (Eqs. \ref{apcosc} and
\ref{apsinc}) 
into:
\begin{equation}
\label{apmodd}
|\vec{r}-\vec{r}_0|_{\pm, f=0}  = 
  \sqrt{(r [\cos\varphi]_{f=0} \cos i - r_0 \cos\varphi_0)^2 +
  (r [\sin\varphi]_{\pm,f=0}  - r_0 \sin\varphi_0)^2},     
\end{equation}
and $r_{\rm lim}$ is defined by the condition:
\begin{equation}
v(r_{\rm lim}) = v_0 \frac{|x|} {\sin i} .
\end{equation}

\clearpage

\begin{figure}
\figurenum{2}
\plotone{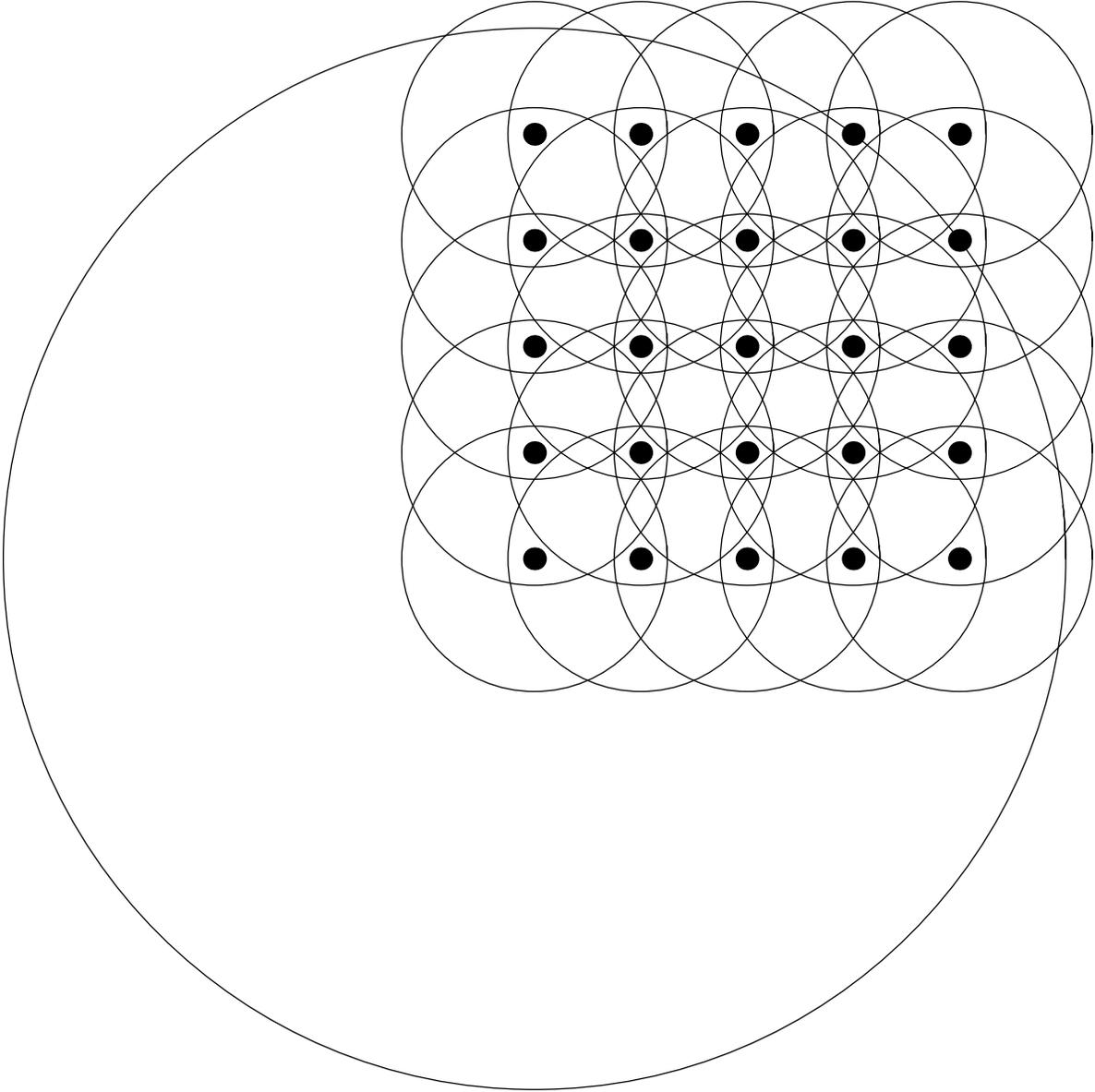}
\figcaption{Grid of relative displacement between the microlens and the BLR. The big
disc represents
the BLR. The small discs correspond to the Einstein circle associated with the
microlens, represented by
a point, in the case
$\eta_0=r_{\rm BLR}/4$. For each point (corresponding
to a displacement of the microlens in the positive quadrant) we compute an
emission-line profile.\label{fig2}}
\end{figure}

\begin{figure}
\figurenum{3}
\plotone{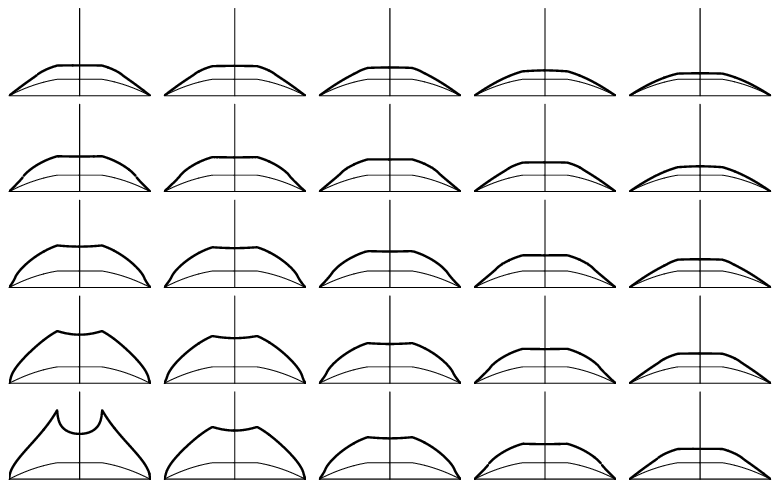}
\figcaption{Spherical model with $p=0.5$, $\beta=-1.5$, and $\eta_0=r_{\rm BLR}$. 
On the $x$-axis we represent $x=c~(\lambda-\lambda_0)/( v_{\rm max}\lambda_0)$,
which varies between $-1$ and 1. On the $y$-axis we represent the flux.
The heavy solid line is the amplified line profile and the solid line is the unamplified 
line profile.\label{fig3}}
\end{figure}

\begin{figure}
\figurenum{4}
\plotone{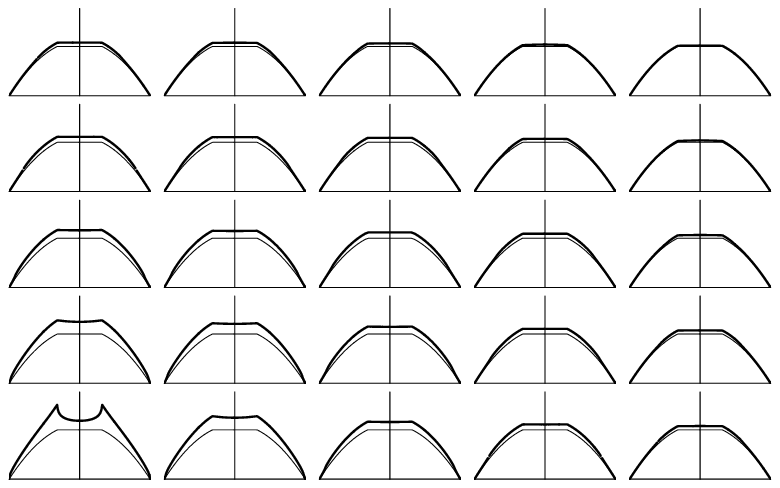}
\figcaption{The same as in Fig. \ref{fig3}, but for $\eta_0=r_{\rm BLR}/4$.\label{fig4}}
\end{figure}

\begin{figure}
\figurenum{5}
\plotone{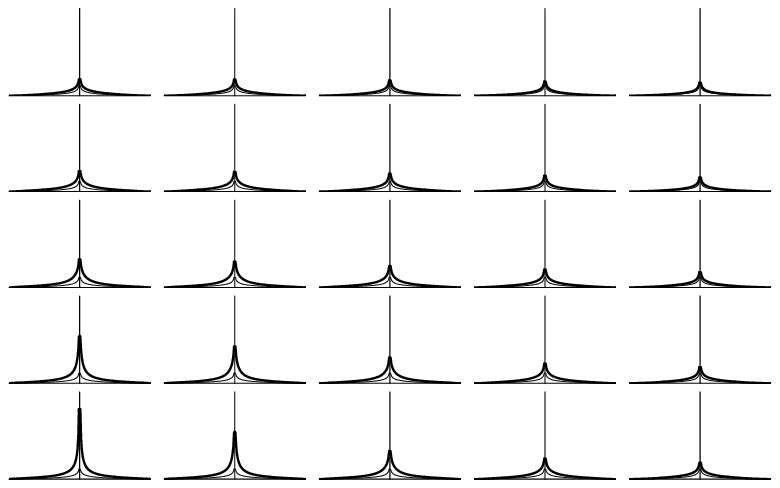}
\figcaption{Spherical model with $p=2$, $\beta=-1.5$, and $\eta_0=r_{\rm BLR}$. 
On the $x$-axis we represent $x=c~(\lambda-\lambda_0)/( v_{\rm max}\lambda_0)$,
which varies between $-1$ and 1. On the $y$-axis we represent the flux.
The heavy solid line is the amplified line profile and the solid line is the unamplified 
line profile.\label{fig5}}
\end{figure}

\begin{figure}
\figurenum{6}
\plotone{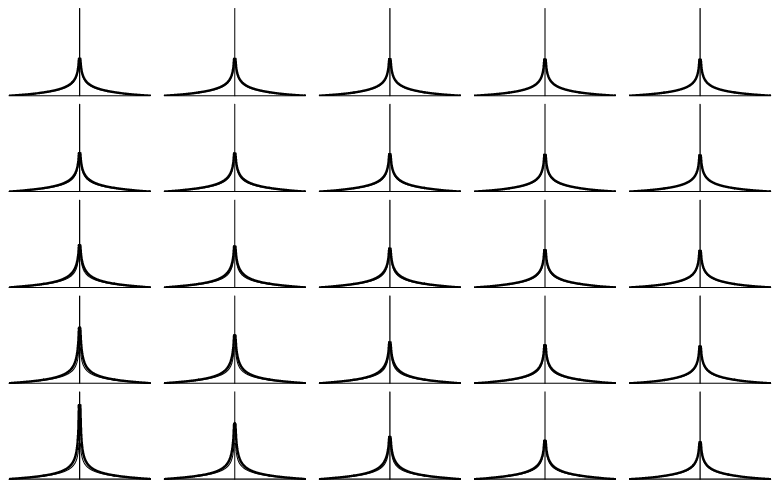}
\figcaption{The same as in Fig. \ref{fig5}, but for $\eta_0=r_{\rm BLR}/4$.\label{fig6}}
\end{figure}

\begin{figure}
\figurenum{7}
\plotone{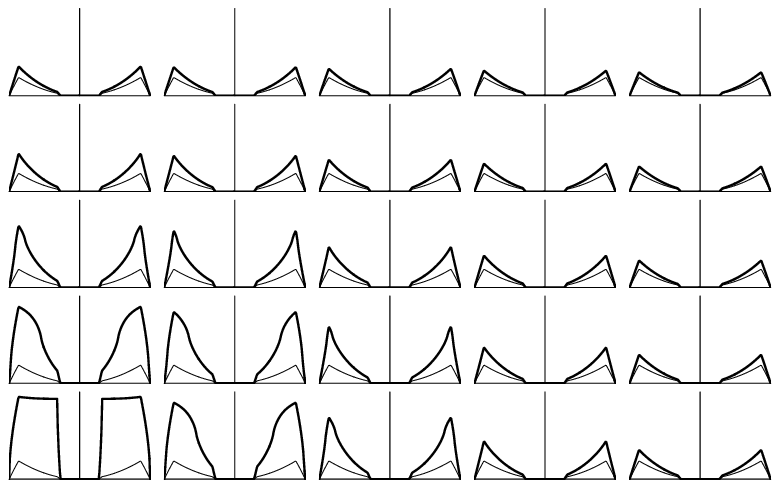}
\figcaption{Biconical model with $i=0^{\circ}$, $p=0.5$, $\beta=-1.5$, and
$\eta_0=r_{\rm BLR}$. On the $x$-axis we represent $x=c~(\lambda-\lambda_0)/( v_{\rm max}\lambda_0)$,
which varies between $-1$ and 1. On the $y$-axis we represent the flux.
The heavy solid line is the amplified line profile and the solid line is the unamplified 
line profile.\label{fig7}}
\end{figure}

\begin{figure}
\figurenum{8}
\plotone{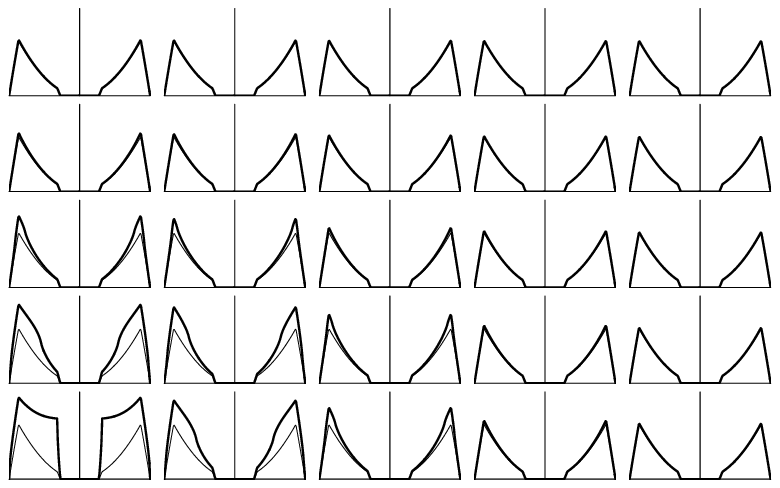}
\figcaption{The same as in Fig. \ref{fig7}, but for $\eta_0=r_{\rm BLR}/4$.\label{fig8}}
\end{figure}

\begin{figure}
\figurenum{9}
\plotone{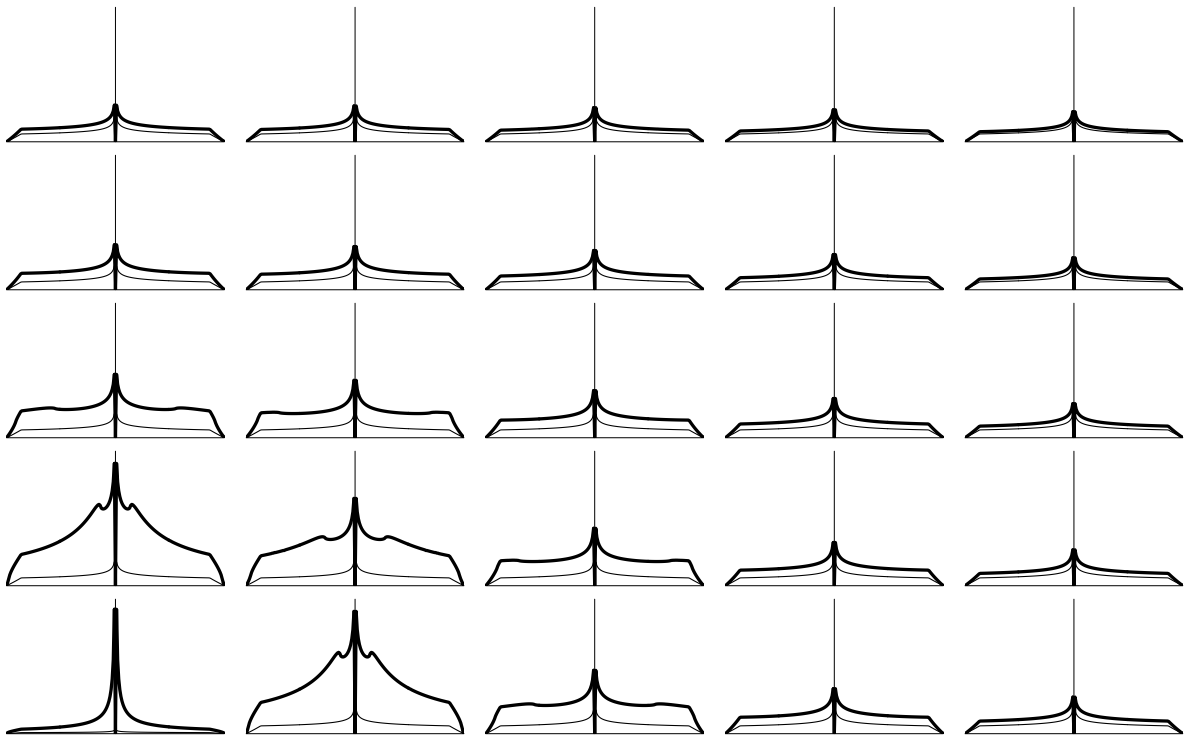}
\caption{Biconical model with $i=0^{\circ}$, $p=2$, $\beta=-1.5$, and $\eta_0=r_{\rm
BLR}$. On the $x$-axis we represent $x=c~(\lambda-\lambda_0)/( v_{\rm max}\lambda_0)$,
which varies between $-1$ and 1. On the $ y$-axis we represent the flux.
The heavy solid line is the amplified line profile and the solid line is the unamplified 
line profile.
The figure in the bottom-left corner has been multiplied by a factor 7.2 for display
purposes.\label{fig9}}
\end{figure}

\begin{figure}
\figurenum{10}
\plotone{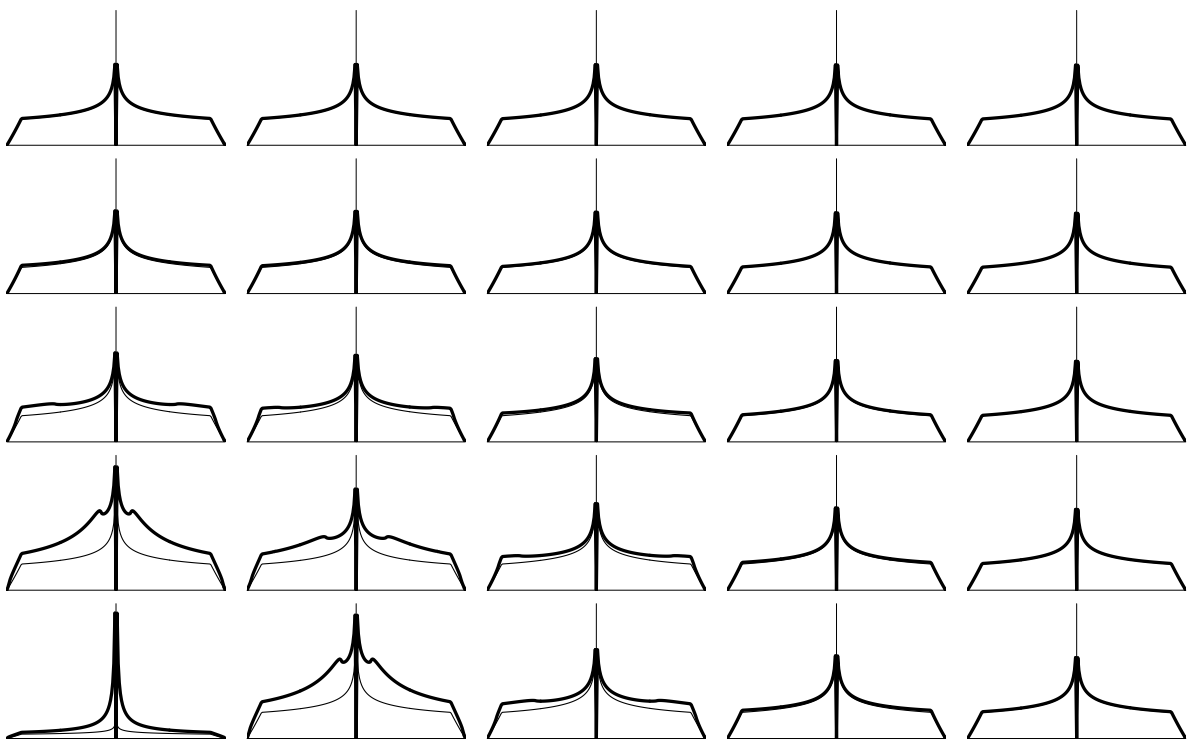}
\caption{The same as in Fig. \ref{fig9}, but for $\eta_0=r_{\rm BLR}/4$.
The figure in the bottom-left corner has been multiplied by a factor 6.0 for display
purposes.\label{fig10}}
\end{figure}

\begin{figure}
\figurenum{11}
\plotone{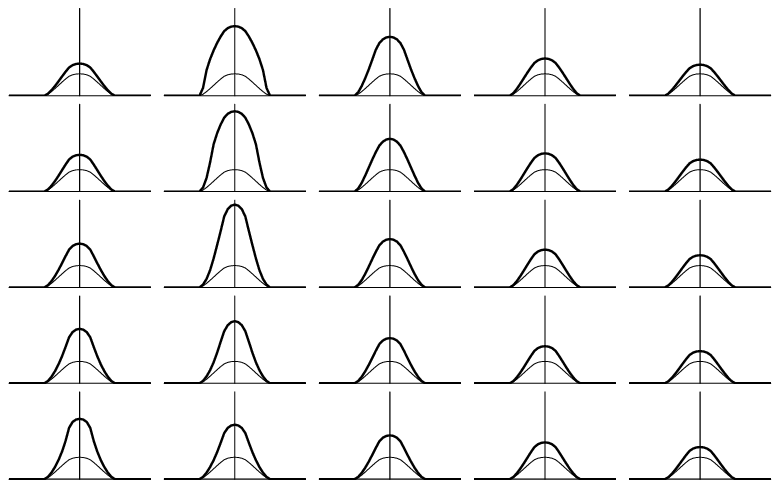}
\caption{Biconical model with $i=90^{\circ}$, $p=0.5$, $\beta=-1.5$, and
$\eta_0=r_{\rm BLR}$. 
On the $x$-axis we represent $x=c~(\lambda-\lambda_0)/( v_{\rm max}\lambda_0)$,
which varies between $-1$ and 1. On the $y$-axis we represent the flux.
The heavy solid line is the amplified line profile and the solid line is the unamplified 
line profile. \label{fig11}}
\end{figure}

\begin{figure}
\figurenum{12}
\plotone{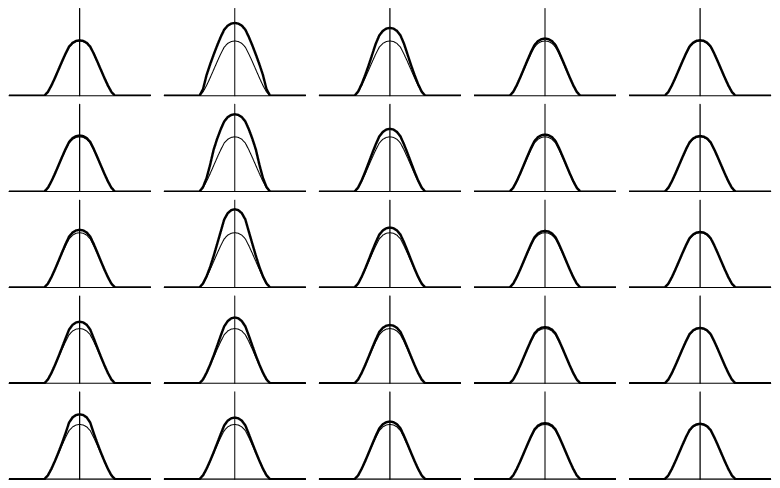}
\caption{The same as in Fig. \ref{fig11}, but for $\eta_0=r_{\rm BLR}/4$.\label{fig12}}
\end{figure}

\begin{figure}
\figurenum{13}
\plotone{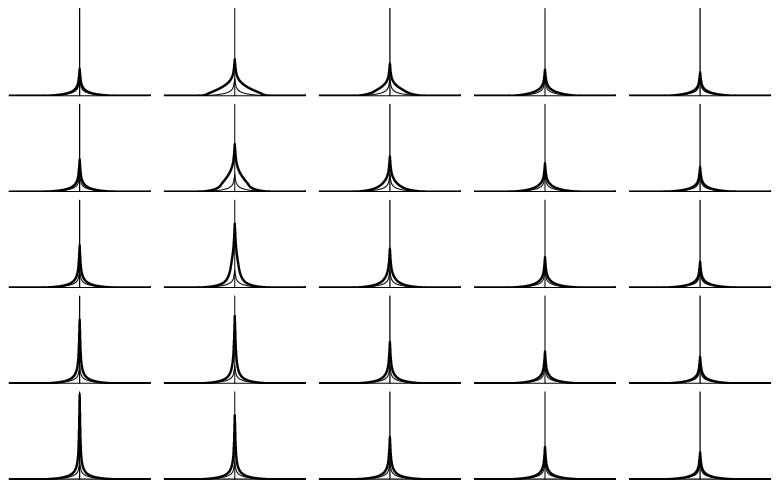}
\caption{Biconical model with $i=90^{\circ}$, $p=2$, $\beta=-1.5$, and
$\eta_0=r_{\rm BLR}$. 
On the $x$-axis we represent $x=c~(\lambda-\lambda_0)/( v_{\rm max}\lambda_0)$,
which varies between $-1$ and 1. On the $y$-axis we represent the flux.
The heavy solid line is the amplified line profile and the solid line is the unamplified 
line profile.\label{fig13}}
\end{figure}

\begin{figure}
\figurenum{14}
\plotone{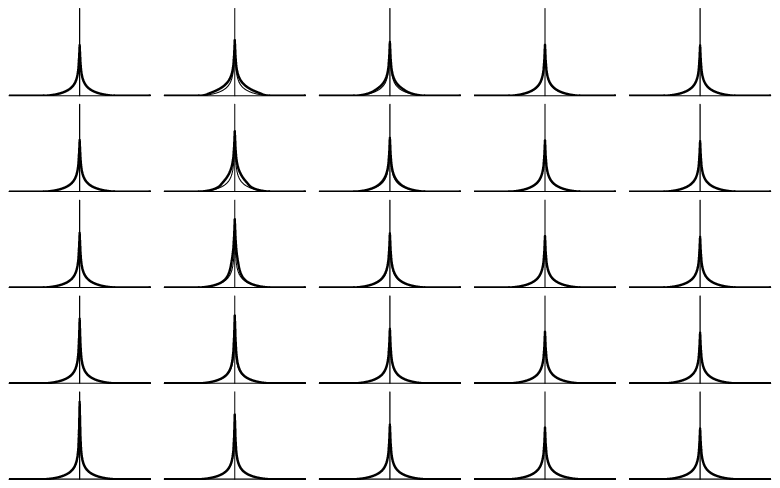}
\caption{The same as in Fig. \ref{fig13}, but for $\eta_0=r_{\rm BLR}/4$.\label{fig14}}
\end{figure}

\clearpage

\begin{figure}
\figurenum{15}
\plotone{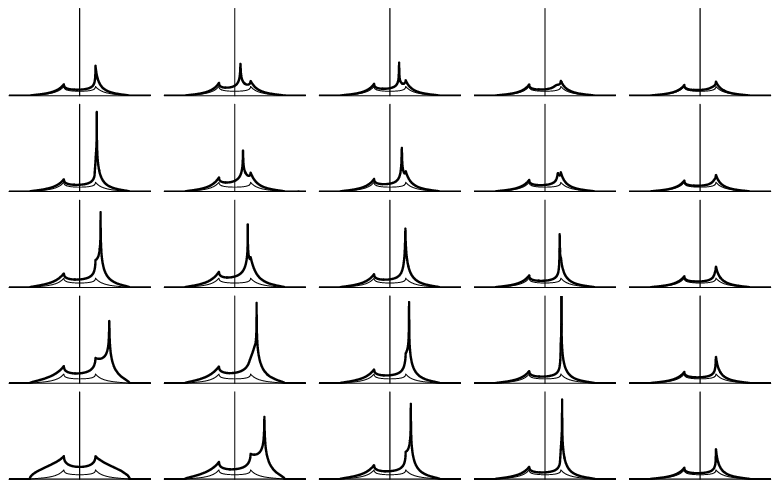}
\caption{Model of Keplerian disk with $i=45^{\circ}$, $p=-0.5$, $\beta=-1.5$, and
$\eta_0=r_{\rm BLR}$. 
On the $x$-axis we represent $x=c~(\lambda-\lambda_0)/( v_{\rm max}\lambda_0)$,
which varies between $-1$ and 1. On the $y$-axis we represent the flux.
The heavy solid line is the amplified line profile and the solid line is the unamplified 
line profile. \label{fig15}}
\end{figure}

\begin{figure}
\figurenum{16}
\plotone{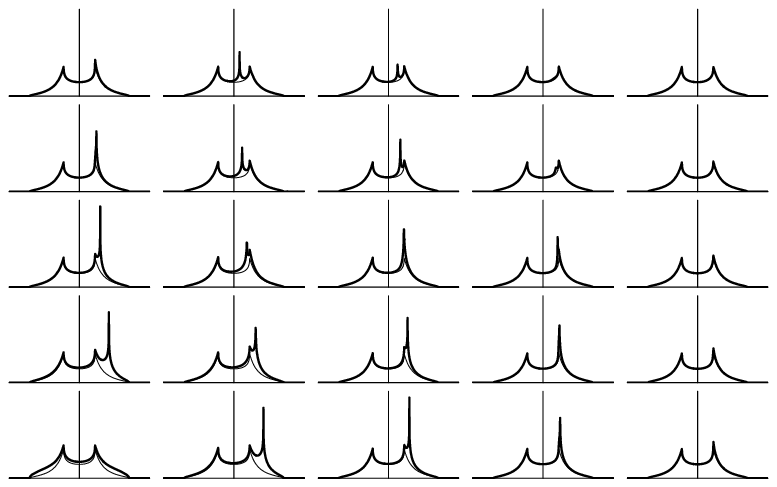}
\caption{The same as in Fig. \ref{fig15}, but for $\eta_0=r_{\rm BLR}/4$.\label{fig16}}
\end{figure}

\begin{figure}
\figurenum{17}
\plotone{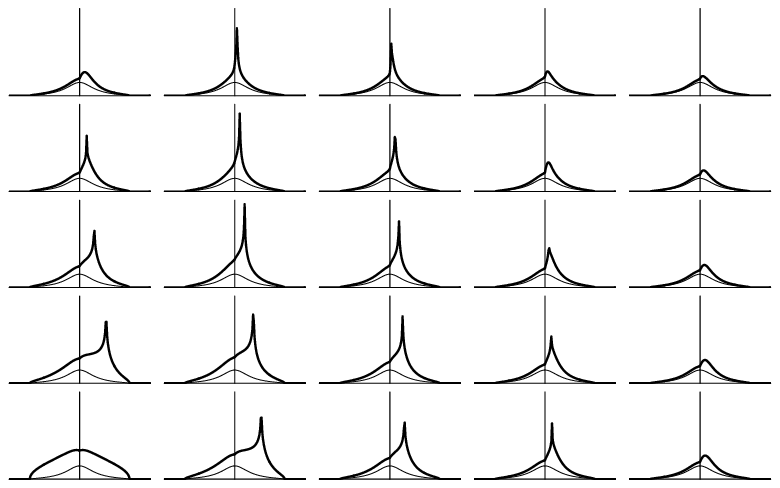}
\caption{Model of modified Keplerian disk with $i=45^{\circ}$, $p=-0.5$,
$\beta=-1.5$, and $\eta_0=r_{\rm
BLR}$. 
On the $x$-axis we represent $x=c~(\lambda-\lambda_0)/( v_{\rm max}\lambda_0)$,
which varies between $-1$ and 1. On the $y$-axis we represent the flux.
The heavy solid line is the amplified line profile and the solid line is the unamplified 
line profile. \label{fig17}}
\end{figure}

\begin{figure}
\figurenum{18}
\plotone{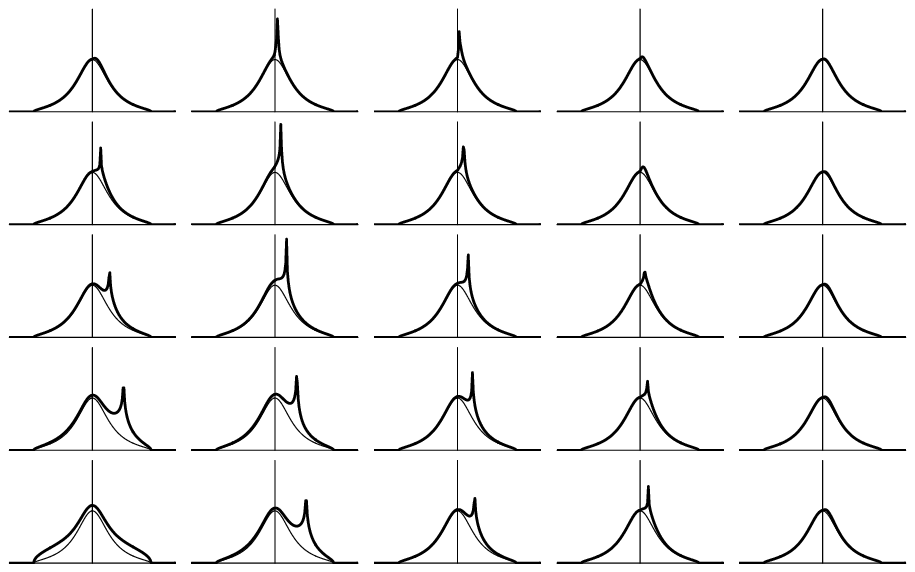}
\caption{The same as in Fig. \ref{fig17}, but for $\eta_0=r_{\rm BLR}/4$.\label{fig18}}
\end{figure}

\begin{figure}
\figurenum{19}
\plotone{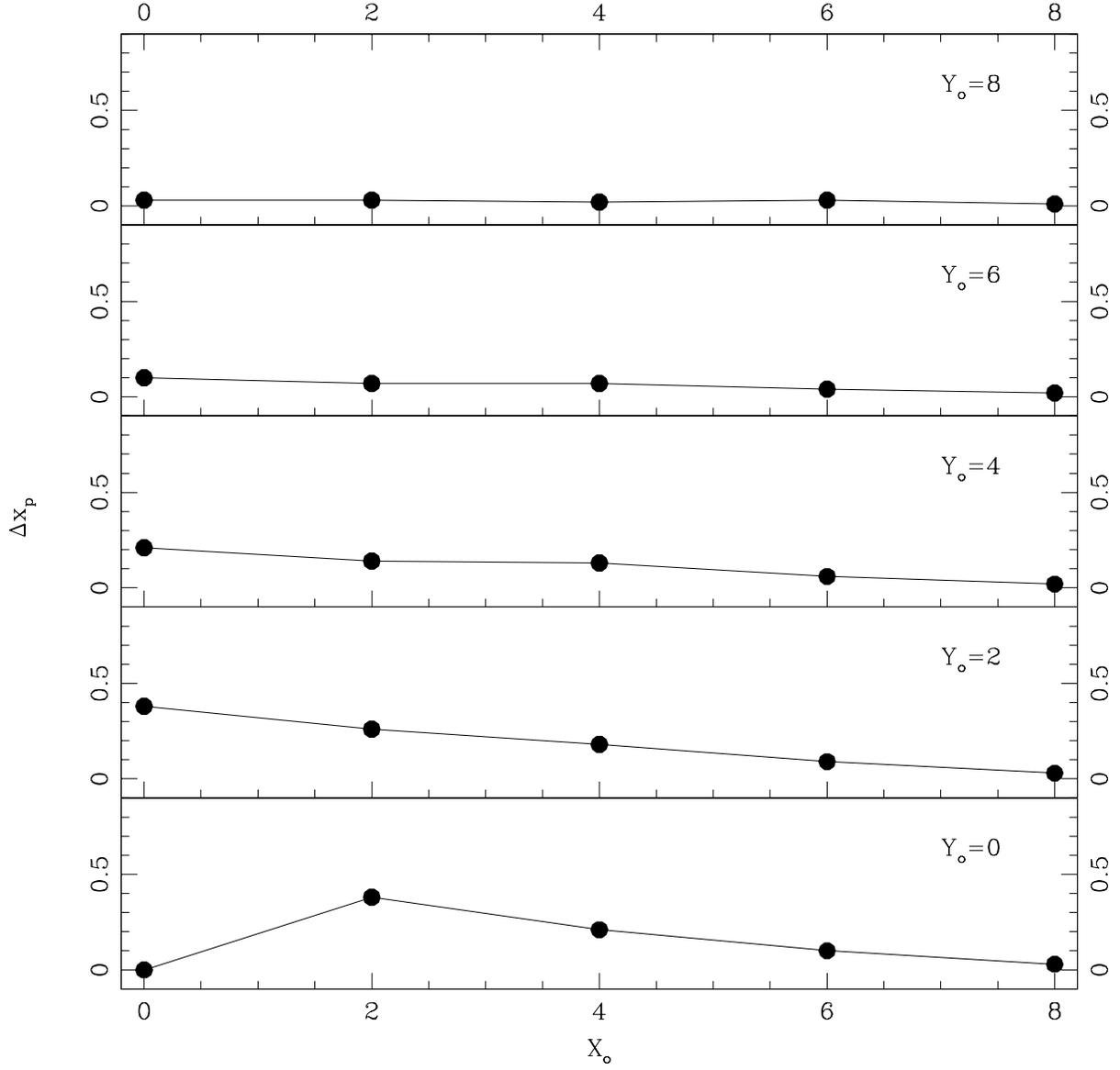}
\caption{Displacements of the line peaks, $\Delta x_{\rm p}$, in the modified
Keplerian disc model with
$i=45^{\circ}$, 
$p=-0.5$, $\beta=-1.5$, and $\eta_0=r_{\rm BLR}/4$ in the different positions
$(X_0,Y_0)$.\label{fig19} }
\end{figure}

\clearpage

\begin{figure}
\figurenum{20}
\plotone{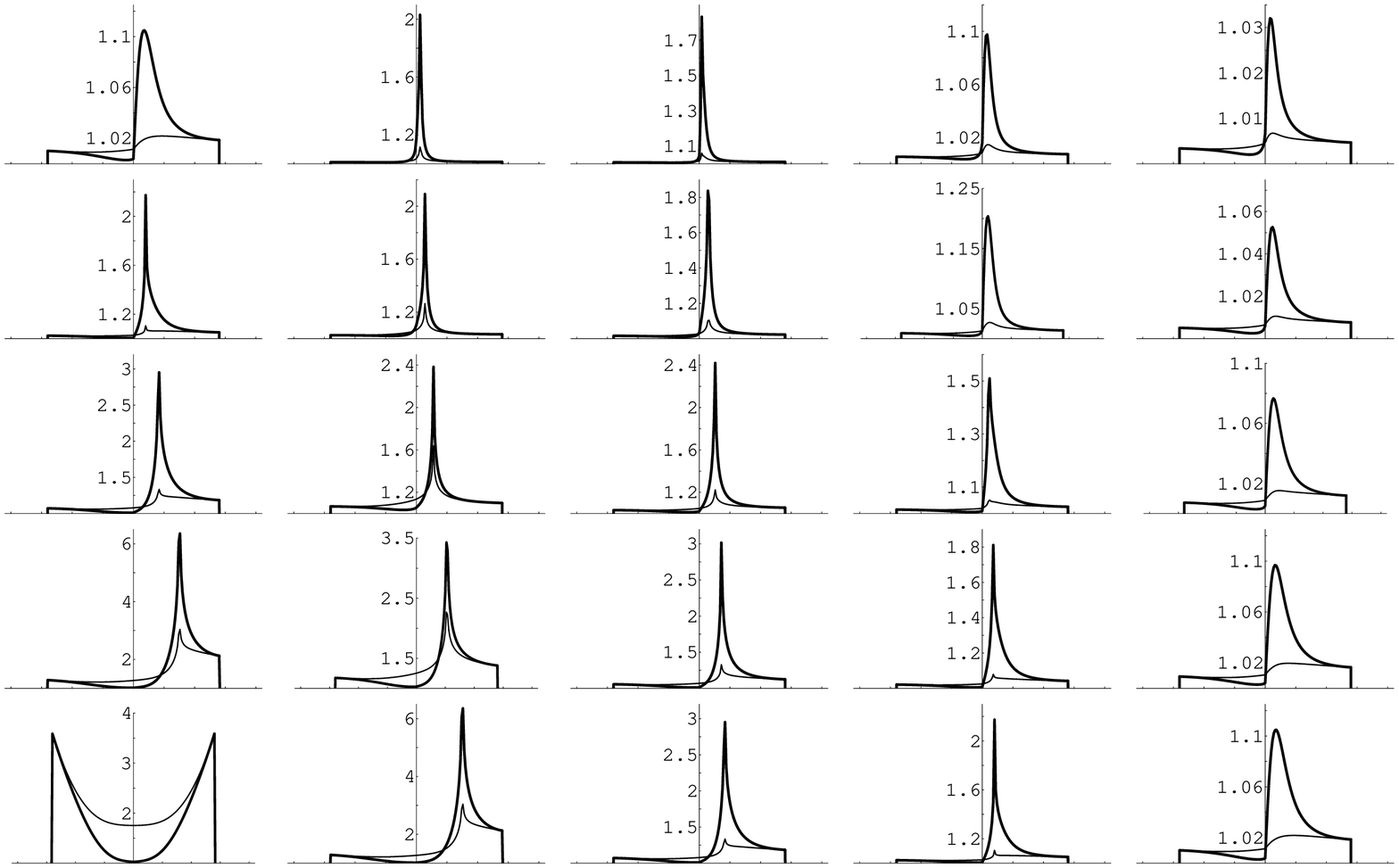}
\caption{Model of modified Keplerian disk with $i=45^{\circ}$, $p=-0.5$, and
$\eta_0=r_{\rm BLR}/4$.
On the $x$-axis we represent $x=c~(\lambda-\lambda_0)/( v_{\rm max}\lambda_0)$,
which varies between $-1$ and 1. On the $y$-axis we represent the ratio between the
amplified flux and the
unamplified flux of the line profile.
The heavy solid line is for $\beta=0$ and the solid line is for $\beta=-4$.\label{fig20}}
\end{figure}


\clearpage

\begin{thebibliography}{}
\bibitem[]{} Alcock, C.  et al. 2000a, \apj, 541, 734
\bibitem[]{} Alcock, C.  et al. 2000b, \apj, 542, 281
\bibitem[]{} Falco, E.~E., Kochanek, C.~S., \& Mu\~noz, J.~A.\ 1998, \apj, 494, 47
\bibitem[]{} Falco, E.~E.~et al.\ 1999, \apj, 523, 617
\bibitem[]{} Hewitt, J.~N., Turner, E.~L., Lawrence, C.~R., Schneider, D.~P., \&
Brody, J.~P.\ 1992, \aj, 104, 968
\bibitem[]{} Jackson, N.~et al.\ 1995, \mnras, 274, L25
\bibitem[]{} Kaspi, S.~et al.\ 2000, \apj, 533, 631
\bibitem[]{} Koratkar, A. P., \& Gaskell, C. M. 1991, \apjs, 75, 719
\bibitem[]{} Leh{\' a}r, J.~et al.\ 2000, \apj, 536, 584
\bibitem[]{} Marziani, P., Calvani, M., \& Sulentic, J. W. 1992, \apj, 393, 658
\bibitem[]{} Mathews, W. G. 1982, \apj, 258, 425
\bibitem[]{} Mediavilla, E.~et al.  1998, \apjl, 503, L27
\bibitem[]{} Mu{\~ n}oz, J.~A.~et al.\ 2001, \apjl, 563, L107
\bibitem[]{} Nemiroff, R. 1988, \apj, 335, 593
\bibitem[]{} Peterson, B. M., \& Wandel, A. 1999, \apjl, 521, L95
\bibitem[]{} Rees, M. J.  1984, \araa, 22, 471
\bibitem[]{} Richards, G.~T.~et al.\ 2001, \aj, 121, 2308
\bibitem[]{} Robinson, A.  1995, \mnras, 272, 647
\bibitem[]{} Schneider, P., Ehlers, J., \& Falco, E. 1992,  Gravitational Lenses (Berlin:
Springer)
\bibitem[]{} Schneider, P., \& Wambsganss, J. 1990, \araa, 273, 42
\bibitem[]{} Wandel, A., Peterson, B. M., \& Malkan, M. A. 1999, \apj, 526, 579
\bibitem[]{} Wyithe, J. S. B., Webster, R. L., \& Turner, E. L. 2000, \mnras, 315, 51
\bibitem[]{} Yonehara, A., Mineshige, S., Fukue, J., Umemura, M., \& Turner, E. L.
1999, \aap, 343, 41
\bibitem[]{} Zheng, W., Binette, M., \& Sulentic, J. W. 1990, \apj, 365, 115
\end{thebibliography}
\end{document}